\documentclass[twocolumn,amsmath,amssymb,superscriptaddress,prb]{revtex4}

\usepackage{subscript}
\usepackage{graphicx}
\usepackage{multirow}

\usepackage[breaklinks]{hyperref}
\hypersetup{
    bookmarks=false,
    pdfstartview={FitH},
    colorlinks=true,
    linkcolor=blue,
    citecolor=blue,
    urlcolor=blue
}

\begin{document}

\title{Enhanced electron correlations at the Sr$_x$Ca$_{1-x}$VO$_3$ surface}

\author{J.~Laverock}
\author{J.~Kuyyalil}
\author{B.~Chen}
\affiliation{Department of Physics, Boston University, 590 Commonwealth Avenue,
Boston, MA 02215, USA}

\author{R.~P.~Singh\footnote{Present address:
Department of Physics, IISER Bhopal, MP-462023, India}}
\affiliation{Department of Physics, University of Warwick, Coventry, CV4 7AL,
United Kingdom}

\author{B.~Karlin}
\author{J.~C.~Woicik}
\affiliation{Materials Science and Engineering Laboratory, National Institute of
Standards and Technology, Gaithersburg, MD 20899, USA}

\author{G.~Balakrishnan}
\affiliation{Department of Physics, University of Warwick, Coventry, CV4 7AL,
United Kingdom}

\author{K.~E.~Smith}
\affiliation{Department of Physics, Boston University, 590 Commonwealth Avenue,
Boston, MA 02215, USA}
\affiliation{School of Chemical Sciences and MacDiarmid Institute for
Advanced Materials and Nanotechnology, University of Auckland,
Auckland 1142, New Zealand}

\begin{abstract}
We report hard x-ray photoemission spectroscopy measurements of the electronic
structure of the prototypical
correlated oxide Sr$_x$Ca$_{1-x}$VO$_3$. By comparing spectra recorded at
different excitation energies, we show that 2.2~keV photoelectrons contain
a substantial surface component, whereas 4.2~keV photoelectrons originate
essentially from the bulk
of the sample. Bulk sensitive measurements of the O~$2p$ valence band are found
to be in good agreement
with {\em ab initio} calculations of the electronic structure with some modest
adjustments to the orbital-dependent photoionization cross-sections. The
evolution of the O~$2p$ electronic structure as a function of the Sr content
is dominated by $A$-site hybridization. Near the Fermi level,
the correlated V~$3d$ Hubbard bands are found to evolve in both binding
energy and spectral weight as a function of distance from the vacuum interface,
revealing higher correlation at the surface than in the bulk.
\end{abstract}

\maketitle

\section{Introduction}
The broad family of complex correlated oxides are the next
frontier in novel materials, providing a rich variety of tunable properties that
are both functionally useful and valuable in understanding fundamental condensed
matter physics. For example, interfaces between perovskite-type oxides are
relatively easily achieved at an atomically precise scale, yielding new and
exciting properties.\cite{chakhalian2012etc} As these materials become
more important, achieving a firm understanding of their surfaces, including the
role and degree that electron correlations play, is crucial.

In strongly correlated materials, the on-site Coulomb energy, $U$, leads to
dynamical correlations between electrons, tending towards their localization at
atomic sites. Incoherent electron states, the Hubbard subbands, form either
side of the coherent, one electron-like quasiparticle peak (QP), and are
referred to as lower and upper Hubbard bands (LHB and UHB respectively).
Transfer of spectral weight away from the QP into the LHB and UHB signals the
effects of stronger electron correlations. Such strong correlations can have
profound effects on the electronic structure and material properties, most
famously leading to insulating, rather than metallic, ground
states.\cite{kotliar2006}

Sr$_x$Ca$_{1-x}$VO$_3$ are prototypical strongly-correlated perovskites, with
similar spectral weight in both the incoherent (strongly correlated) Hubbard
subbands and the coherent (one electron-like) quasiparticle states. Although the
spectral function of these materials has been well studied experimentally
\cite{laverock2013c,sekiyama2004,yoshida2005etc,eguchi2006,yoshida2010,
maiti2001,maiti2006,pen1999etc,mossanek2008,takizawa2009} and
theoretically,\cite{liebsch2003,nekrasov2005,ishida2006,nekrasov2006,
byczuk2007etc} there is not a well established consensus on the form of
the surface and bulk components, or as the Sr content ($x$) is varied.
Both endmembers are correlated metals and have been reported to exhibit
metal-insulator transitions in ultrathin films due to a reduction in the
dimensionality and corresponding narrowing of the V~$3d$
bandwidth.\cite{yoshimatsu2010etc}
Generally, the focus of
photoemission spectroscopy (PES) studies has been on obtaining
bulk-representative spectra. For example, laser-PES has been used at very low
photon energies, below the minimum in the photoelectron inelastic mean free path
(IMFP), to extract bulk QP spectra,\cite{eguchi2006} although the LHB is not
accessible at these low energies. Conversely, soft x-rays have also been used to
increase the depth sensitivity, and PES spectra were obtained which were either
independent of $x$\cite{sekiyama2004} or which revealed an insulating
surface.\cite{maiti2001} More
recently, some progress has been made in understanding the
behavior of the spectra with $x$, with several reports agreeing that the LHB of
CaVO$_3$ is more intense and closer to the Fermi level, $E_{\rm F}$, than that
of SrVO$_3$,\cite{maiti2006,yoshida2010} indicative of the effects of stronger
electron correlations. Additionally, band dispersions and Fermi surfaces
have been clearly observed in both
end-members.\cite{yoshida2005etc,takizawa2009,yoshida2010} However, the
behavior of the spectra at the surface compared with the bulk has received
little experimental attention, despite the role of the surface being
questioned and studied in detail theoretically.\cite{liebsch2003,ishida2006}
Moreover, our knowledge of the ``bulk'' spectra from PES has so far been
limited to analytical differences between spectra containing both bulk and
surface components.\cite{eguchi2006,sekiyama2004,maiti2001,maiti2006}

Recently, resonant soft x-ray emission spectroscopy (RXES) measurements have
reported truly bulk-representative information on the strongly-correlated V~$3d$
bands, illustrating that the effects of electron correlations are more
pronounced in CaVO$_3$ than in Sr$_{0.5}$Ca$_{0.5}$VO$_3$, and that the surfaces
of both materials are ``more correlated'' than their bulk.\cite{laverock2013c}
Although valuable, comparisons between RXES and PES spectra are complicated by
the quite different scattering processes and comparatively poorer resolution of
RXES (in part, owing to the broadening due to the lifetime of the core hole). In
order to elicit quantitative information on the evolution of electron
correlations from the surface to the bulk, depth sensitive measurements using
the same probe are clearly desired.

Hard x-ray photoemission spectroscopy (HAXPES) extends conventional PES
measurements at ultraviolet and soft x-ray energies
into the hard x-ray regime ($\gtrsim 4$~keV), yielding much
greater depth sensitivity ($\gtrsim 5$~nm compared with 0.5~nm at ultraviolet
energies).\cite{fadley2005} However, at such hard energies, there are several
additional considerations. Most importantly, the cross section for
photoionization drops off at higher energies, particularly for shallow binding
energies at the valence band, \cite{trzhaskovskaya2001}
and practical instrument resolutions are
poorer than at low energies. Additionally, the electron momentum distribution is
focused into much tighter angles, meaning the coverage of the Brillouin zone
of a typical HAXPES measurement is much broader, although
recent technical advances have facilitated the momentum-resolved spectral
function at hard x-ray energies.\cite{gray2011}

We report here a HAXPES study of the electronic structure of
Sr$_x$Ca$_{1-x}$VO$_3$ using photon energies of 4.2~keV and 2.2~keV. The
valence band of Sr$_x$Ca$_{1-x}$VO$_3$ measured using
4.2~keV photons is quantitatively compared with {\em ab
initio} theoretical calculations of the bulk, and we show that the spectra can
be well described by band theory with some modest adjustments to the
orbital-dependent photoionization cross-sections. Additionally, we find that
the evolution of the valence band electronic structure with $x$
is dominated by different $A$-site hybridization.
Secondly, the electronic structure of the correlated
V~$3d$ states is investigated as a function of distance into the sample from the
surface. This was achieved by varying the photon
energy of the HAXPES measurements, and by comparison with previous
surface-sensitive ultraviolet photoemission spectroscopy (UPS) and unambiguously
bulk-sensitive RXES measurements.
Particular attention was paid to the {\em energetics} of the Hubbard
subbands, which are one of the key properties in identifying the evolution
of correlated behavior. We find that spectra recorded at 2.2~keV excitation
energy still contain a large surface
contribution of 30 -- 40\%, despite photoelectrons at this energy often being
considered bulk probes of correlated materials. However, we show that
photoelectrons at 4.2~keV are found to essentially probe the bulk electronic
structure. Significantly,
we demonstrate that the correlated localized Hubbard bands evolve in
both energy and spectral weight from the surface (``more correlated'') of the
material to the bulk (``less correlated''), in remarkable agreement with
dynamical mean-field theory predictions.

\begin{figure}[t!]
\begin{center}
\includegraphics[width=0.8\linewidth,clip]{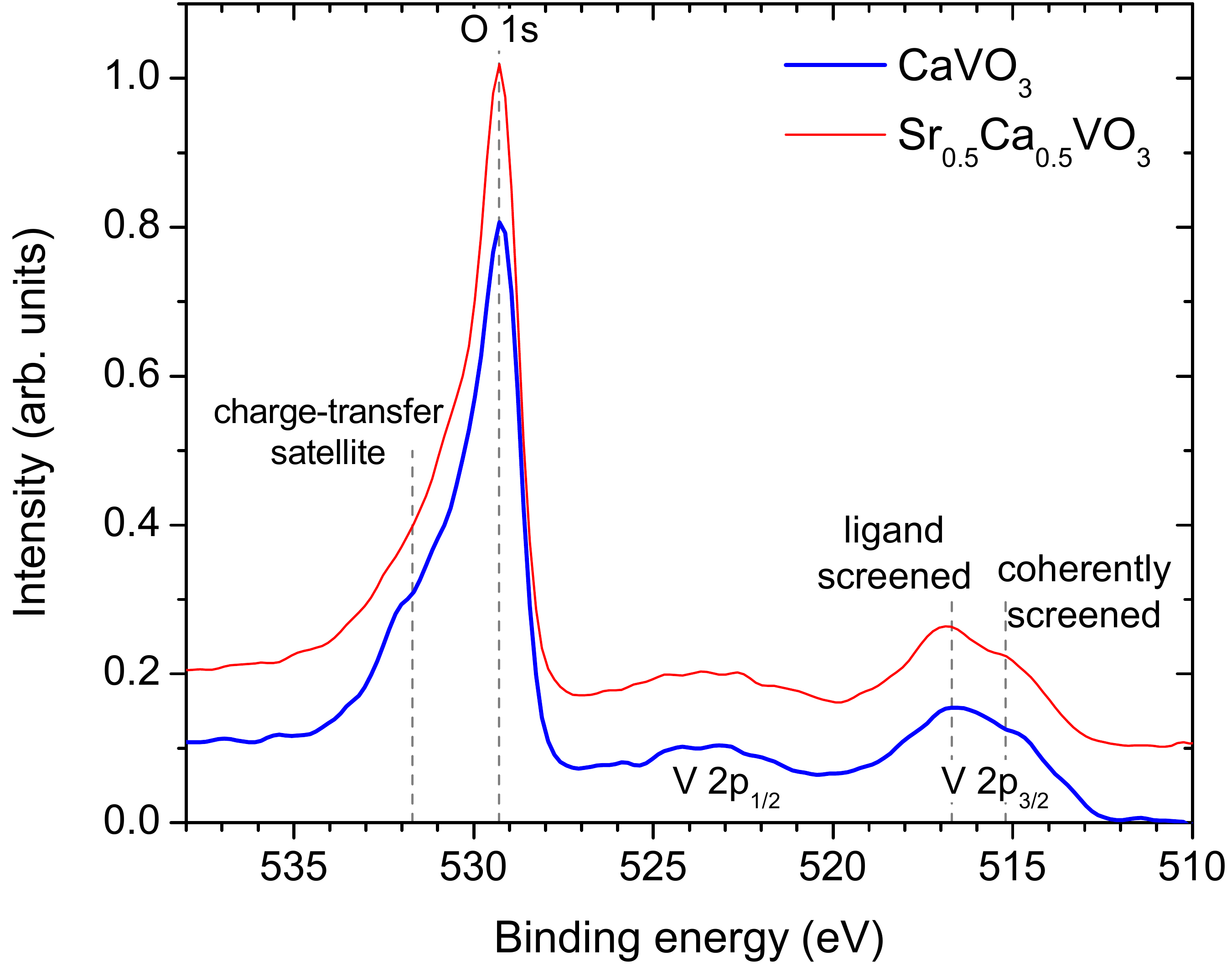}
\end{center}
\vspace*{-0.2in}
\caption{\label{f:shcore} (Color online) Core level HAXPES spectra at 4.2~keV of
CaVO$_3$ and Sr$_{0.5}$Ca$_{0.5}$VO$_3$.}
\end{figure}

\section{Methods}
Large high quality single crystals of CaVO$_3$ (CVO) and
Sr$_{0.5}$Ca$_{0.5}$VO$_3$ (SCVO) were grown by the floating zone technique in a
four mirror optical furnace, employing growth rates of 7 to 10~mm/h in an
atmosphere of 1~bar of Ar + 3\% H$_2$ gas.\cite{laverock2013c,inoue1998}
HAXPES measurements
were performed at Beamline X24A of the National Synchrotron Light Source,
Brookhaven National Laboratory, with total instrument resolution ($\sigma$)
of 180~meV at
2.2~keV and 220~meV at 4.2~keV. Photoelectrons with kinetic energies of 2.2~keV
and 4.2~keV are estimated to have an IMFP, $\lambda$,
of 3.6~nm and 6.1~nm respectively in Sr$_x$Ca$_{1-x}$VO$_3$ within the
TPP-2M model,\cite{tanuma1994} compared with 0.5~nm at ultraviolet
energies ($\sim 80$~eV).\cite{laverock2013c,yoshida2010} Owing to the
different densities of CaVO$_3$ and SrVO$_3$, the IMFP within this model
is anticipated to be $\sim 8$\% smaller in SrVO$_3$ than in CaVO$_3$ (mean
values, roughly corresponding to Sr$_{0.5}$Ca$_{0.5}$VO$_3$, are quoted above).
Samples were cleaved {\em ex-situ}, and
immediately loaded into the ultra-high vacuum chamber. The photoelectron energy
was referenced to polycrystalline Ag in electrical contact with the sample.

HAXPES measurements at 4.2~keV of the O~$1s$ and V~$2p$ core levels are shown in
Fig.~\ref{f:shcore}, and are in good agreement with previous reports at
1.8~keV.\cite{mossanek2008} The two compounds exhibit essentially identical
core level spectra. The V~$2p_{3/2}$ level exhibits two structures,
originating from the ligand-screened and coherently-screened final states of the
photoemission process, and the $2p_{3/2}$ charge-transfer satellite is observed
as a shoulder to the O~$1s$ peak. These spectra are free of surface contaminant
species and disorder effects, such as reports of V$^{3+}$/V$^{5+}$
segregation at the surface of scraped samples.\cite{maiti2001,maiti2006}
We emphasize that the enhanced electron correlations at the surface that we
discuss in Section~\ref{s:nearef} are an {\em intrinsic} property of the
surface, arising from the broken translational symmetry and reduced
coordination,\cite{ishida2006} rather than an extrinsic
surface problem (due to contaminant species and/or surface disorder).

{\em Ab initio} electronic structure calculations have been performed using the
all-electron full-potential linearized augmented plane-wave (FLAPW) implemented
by the {\sc Elk} code,\cite{elk} within the local density
approximation (LDA) to the exchange-correlation functional.
Experimental lattice parameters of 3.84~{\AA} was used for
cubic SrVO$_3$\cite{inoue1998} and $a = 5.32$~{\AA}, $b = 5.34$~{\AA} and $c =
7.55$~{\AA} were used for orthorhombic CaVO$_3$.\cite{falcon2004}
Convergence was achieved over a $13 \times 13 \times 13$ mesh in
the full Brillouin zone.

\begin{figure}[t!]
\begin{center}
\includegraphics[width=1.0\linewidth,clip]{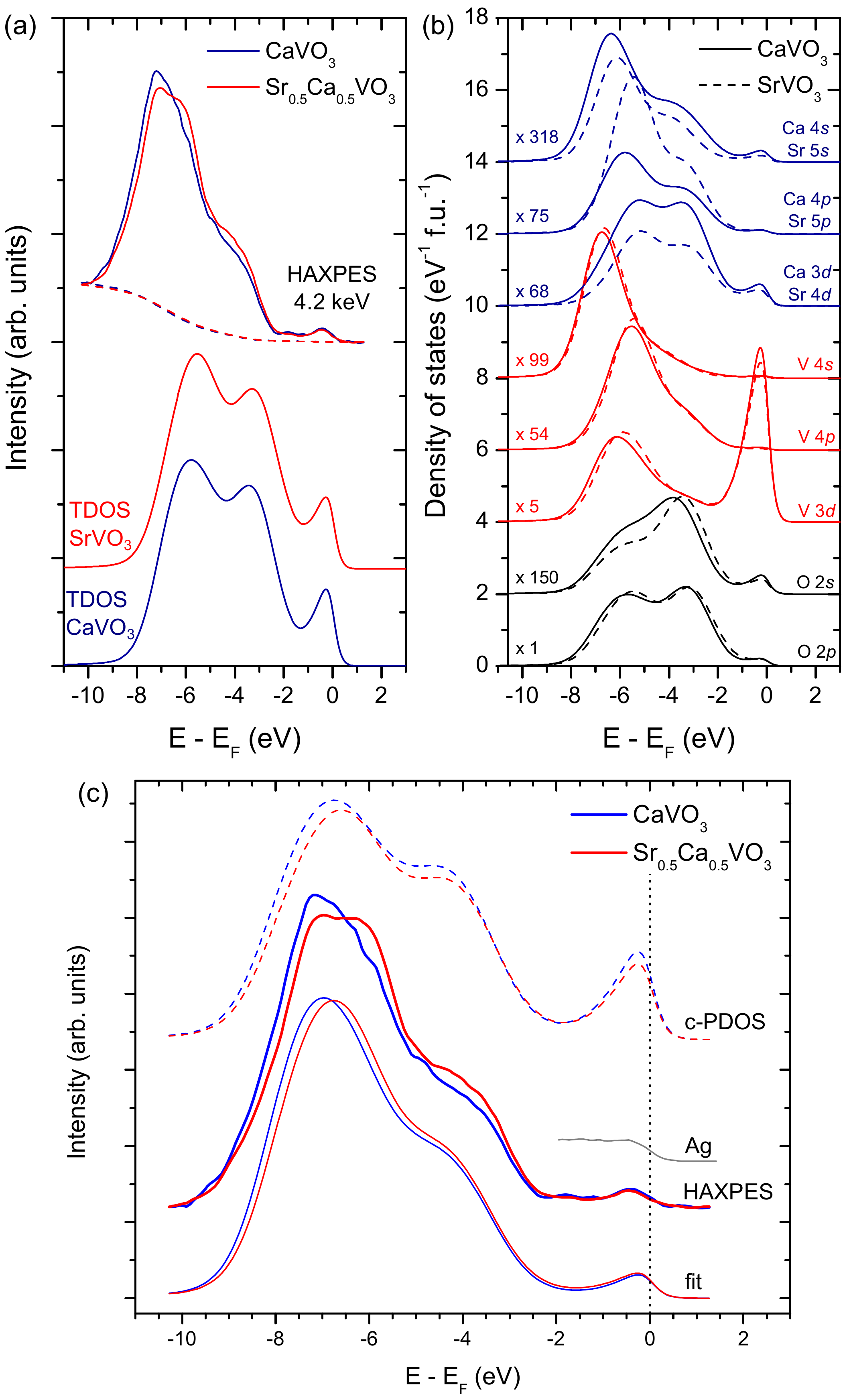}
\end{center}
\vspace*{-0.2in}
\caption{\label{f:vb2p} (Color online) (a)~HAXPES measurements at
4.2~keV of CaVO$_3$ and Sr$_{0.5}$Ca$_{0.5}$VO$_3$, illustrating the
Shirley-type background contribution (dashed line). The broadened total
DOS (TDOS) is shown at the bottom of the figure for CaVO$_3$ and SrVO$_3$.
(b)~Broadened partial densities of states of CaVO$_3$ and SrVO$_3$, showing
the contribution from each orbital component, and their respective weights. Each
curve has been offset vertically by 2~eV$^{-1}$~f.u.$^{-1}$ for clarity.
(c)~HAXPES measurements after subtraction of the Shirley-type background,
compared with the results of fitting the PDOS to the data (fit). Also shown
is the theoretical cross-section corrected PDOS (c-PDOS).}
\end{figure}

\section{Valence band electronic structure}
\label{s:vb}
HAXPES measurements at 4.2~keV of the O $2p$ and
V $3d$ valence band are shown for CVO and SCVO
in Fig.~\ref{f:vb2p}(a). The O
$2p$ states are characterized by a strong peak at $-7$~eV, with a prominent
shoulder at $-4$~eV, above which the V $3d$ states are weakly visible as a
double-peaked structure (and to which we return in more detail below).  These
spectra are in approximate agreement with previous PES measurements at
low\cite{takizawa2009} and high\cite{mossanek2008} photon energies. For
comparison, the total (broadened) density of states (DOS)
from the FLAPW calculation, neglecting the
energy-dependent photoionization cross-section, is shown for the pure compounds
at the bottom of Fig.~\ref{f:vb2p}(a). In Fig.~\ref{f:vb2p}(b), the broadened
occupied partial densities of states (PDOS)
of CaVO$_3$ and SrVO$_3$ are shown, scaled (by a factor shown in
the figure) to have approximately the same area. Whereas the total DOS are
dominated by O $2p$ and V $3d$ character, the orbital-dependent photoionization
cross-section at $\approx 4$~keV favors photoemission from the more localized
$s$ and $p$ states.  For example, the strength of the peak at $-7$~eV cannot be
explained without a dramatic enhancement in the V $sp$ and Ca/Sr $s$ states,
compared with the O $2p$ PDOS. Indeed, this is reflected in theoretical
tabulations of these cross-sections, which predict the Ca $4s$, Sr $5s$ and V
$4s$ cross-sections are between 13 and 21 times larger than the O $2p$
cross-section at 4~keV.\cite{trzhaskovskaya2001}

In order to more accurately assess the origin of the differences in the HAXPES
spectra between CVO and SCVO, we turn to a more quantitative analysis, in which
the relative photoionization cross-sections for each orbital are fitted to the
experimental data, yielding approximate {\em empirical} relative cross-sections.
The data are first corrected for inelastic scattering processes, leading to the
Shirley-type background shown by the dashed line in Fig.~\ref{f:vb2p}(a).  The
theoretical spectrum is then computed as a cross-section weighted sum of the
orbital PDOS components shown in Fig.~\ref{f:vb2p}(b), neglecting the
interstitial DOS (which, being delocalized away from the ion centers, are
unlikely to contribute significantly at hard x-ray energies), and compared to
the experimental spectrum. For SCVO, we take the rather crude approximation of
50\% CaVO$_3$ PDOS and 50\% SrVO$_3$ PDOS, although in practice this has only a
very weak influence on the results; for example, comparing pure SrVO$_3$ with
the SCVO data yields almost identical results.

The agreement between experiment
and theory is optimized for both SCVO and CVO simultaneously using a single set
of parameters. Additionally, the O $2p$ manifold is allowed to be rigidly
shifted in energy during the fit, accounting for the inadequate description by
the LDA of the V $3d$ bands. In the LDA FLAPW calculations, the LHB feature
(the deeper of the two V $3d$ peaks) is absent, and so this rigid shift
physically represents the experimental separation between the O $2p$ and {\em
coherent} V $3d$ manifolds.  Finally, the experimental broadening of the DOS is
fitted through the convolution of the theoretical DOS with a Voigt function,
$V(\sigma,\gamma) = G(\sigma) \ast L(\gamma)$, where $G(\sigma)$ is a Gaussian
of width $\sigma$ and $L(\gamma)$ is a Lorentzian of width $\gamma$. The results
of the fit yield $\sigma = 0.67$~eV and $\gamma = 0.13$~eV, accounting for the
total broadening due to both intrinsic and experimental resolution effects. All
theoretical spectra in Fig.~\ref{f:vb2p} have been convoluted with this
function, and multiplied by a Fermi function of $k_{\rm B}T_{\rm eff} =
0.145$~eV [from a
separate fit to the Fermi edge of polycrystalline Ag at 4.2~keV shown in
Fig.~\ref{f:vb2p}(c)]. Although the solution from this fitting procedure is
found to be relatively robust and stable, the extent to which the fitted
parameters represent real photoionization cross-sections must be considered with
caution. For example, inaccuracies in the background subtraction and energy
efficiency of the measurement may contribute non-trivially to the fit.
Secondly, inadequacies of the LDA, which may under- or overestimate
hybridization between orbital characters, may be strongly exaggerated, and
the LHB feature is completely absent from the theoretical curve.
Nevertheless, as we discuss below, the results do reflect physical expectations,
and are interpreted as providing a reasonable indication of the trends of the
orbital cross-sections.

\begin{table}[t!]
\begin{center}
\begin{tabular}{|lllr@{.}llr@{.}lc|cllr@{.}llr@{.}ll|}
\hline
\phantom{i} & orbital & \phantom{i} & \multicolumn{5}{c}{relative CS} &
\phantom{i} & \phantom{i} &
orbital & \phantom{i} & \multicolumn{5}{c}{relative CS} & \phantom{i} \\
&&&\multicolumn{2}{l}{theory\cite{trzhaskovskaya2001}}&
&\multicolumn{2}{l}{fitted} &
\phantom{i} & \phantom{i} &
&&\multicolumn{2}{l}{theory\cite{trzhaskovskaya2001}}&
&\multicolumn{2}{l}{fitted} & \\
\hline
& Ca $4s$ && 13&1   && 565&2 &&& V  $4s$ && 20&7   &&  76&7 & \\
& Ca $4p$ && 13&1 * &&  57&3 &&& V  $4p$ && 20&7 * &&  95&5 & \\
& Ca $3d$ &&  0&6 * &&   0&2 &&& V  $3d$ &&  1&7   &&   0&4 & \\
& Sr $5s$ && 16&6   && 737&3 &&& O  $2s$ && 38&5   &&   0&1 & \\
& Sr $5p$ && 16&6 * &&  33&5 &&& O  $2p$ &&  1&0   &&   1&0 & \\
& Sr $4d$ && 10&0 * && 131&9 &&& \multicolumn{7}{c}{}     & \\
\hline
\end{tabular}
\end{center}
\caption{\label{t:fit} Orbital-dependent photoionization cross-sections (CS)
relative to the O $2p$ states according to the theoretical tabulations of
Ref.~\onlinecite{trzhaskovskaya2001} at 4~keV and from the results of fitting
HAXPES spectra at 4.2~keV of CVO and SCVO to the FLAPW PDOS. The theoretical
cross-sections marked by * correspond to nominally empty states that are not
tabulated in Ref.~\onlinecite{trzhaskovskaya2001}. Instead, for the $A$ and V
$p$ states we adopt the corresponding $s$ cross-section, and for the Ca $3d$ and
Sr $4d$ states we use the Sc $3d$ and Y $4d$ cross-sections respectively.}
\end{table}

In Fig.~\ref{f:vb2p}(c), the results of the fit are shown alongside the
background-corrected HAXPES spectra. A shift in the O $2p$ manifold
downwards in energy by $0.9$~eV is required to align the energies of
the peaks in the theoretical and experimental spectra (the same shift is
applied to both CVO and SCVO spectra). The fitted spectrum satisfyingly
reproduces the experimental data, exhibiting the strong peak at $-7$~eV
and its shoulder at $-4$~eV, as well as the rather weak relative intensity
of the coherent V $3d$ states. Moreover, the weak upwards shift in energy of
the lower peak at $-7$~eV of SCVO is captured, along with the increase in the
relative intensity of the shoulder. For comparison, the spectrum due to the
theoretical cross-sections\cite{trzhaskovskaya2001} (labeled c-PDOS)
is also shown (including the rigid shift of the O $2p$ manifold) in
Fig.~\ref{f:vb2p}(c). Although this spectrum roughly reproduces the relative
intensities (and energies) of the O $2p$
manifold, it strongly overestimates the contribution from the V $3d$ electrons.
The fitted, empirical relative cross-sections from the fit are shown in
Table~\ref{t:fit}, and exhibit the expected sensitivity of the HAXPES
experiment to the more localized $s$ and $p$ states. Surprisingly, the relative
O character is strongly suppressed, including the O $2s$ states, in contrast to
other studies at similar energies (which do not attempt to
optimize agreement between
experiment and theory).\cite{woicik2002,woicik2007} However, it is clear from
Fig.~\ref{f:vb2p}(b) that the shape of the O $2s$ PDOS is incompatible with the
data. Finally, we note that incorporating an asymmetric Doniach-\v{S}unji\'{c}
lineshape,\cite{doniach1970} which
has recently been used to explain the HAXPES spectrum of
V$_2$O$_3$,\cite{woicik2007}
does not improve agreement between experiment and theory.

Overall, these results establish that the experimental HAXPES spectra of
CaVO$_3$ and Sr$_{0.5}$Ca$_{0.5}$VO$_3$
can be described well by the theoretical PDOS, with some modest adjustments
to the theoretical orbital cross-sections. Secondly, they demonstrate that the
differences between the experimental spectra, in both relative intensity and
energy, are mostly a consequence of the different $A$-site hybridization. The
importance of $A$-site hybridization in $3d^1$ perovskites has been pointed out
in detail by Ref.~\onlinecite{pavarini2005}. Finally, the effects of the
$A$-site hybridization are more pronounced in the experimental spectra than
predicted by theory.

\section{Near-{\em E}\textsubscript{F} correlated electronic structure}
\label{s:nearef}
\begin{figure}[t!]
\begin{center}
\includegraphics[width=0.9\linewidth,clip]{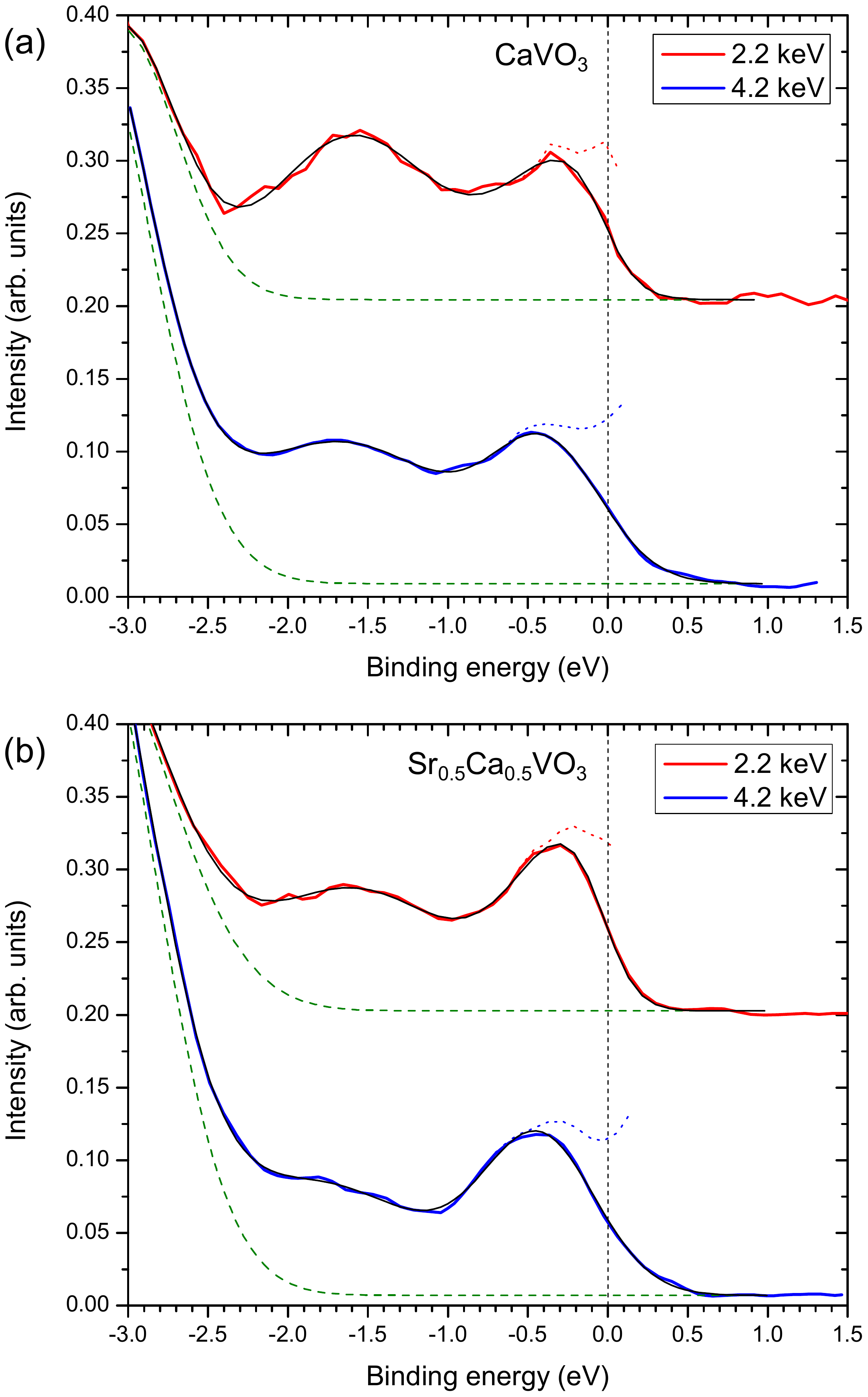}
\end{center}
\vspace*{-0.2in}
\caption{\label{f:vb3d} (Color online) High-statistics HAXPES spectra of the V
$3d$ states of (a)~CaVO$_3$ and (b)~Sr$_{0.5}$Ca$_{0.5}$VO$_3$ at 4.2~keV and
2.2~keV incident photon energies (vertically offset for clarity). The
dashed (green) lines indicate the contributions
from the tails of the O $2p$ band, and the black solid lines represent fits of
the experimental spectra (see text). Finally, the dotted lines near $E_{\rm F}$
are the experimental spectra divided by the Fermi function (determined from Ag
foil).}
\end{figure}

We now address the near-$E_{\rm F}$ electronic structure, which
is dominated by the V $3d$ band. High-statistics spectra
of the V $3d$ manifolds of CVO and SCVO at 4.2~keV and 2.2~keV are shown in
Fig.~\ref{f:vb3d}, referenced to the Fermi level of polycrystalline Ag foil in
electrical contact with the samples. As illustrated by the relatively large
density of states at $E_{\rm F}$, CVO and SCVO are both metallic in the HAXPES
measurement, and consist of the O $2p$ tails below $-2$~eV, followed by the
LHB near $-1.5$~eV and the QP at $E_{\rm F}$. The dashed lines between $-0.5$
and 0~eV represent the same spectra after dividing by the ``effective'' Fermi
function ($k_{\rm B}T_{\rm eff} = 120$~meV
at 2.2~keV and $145$~meV at 4.2~keV), reflecting approximately flat QP DOS
up to $E_{\rm F}$. The ``effective'' Fermi function is determined by fitting a
Fermi function to the polycrystalline Ag reference spectra (displayed below in
Fig.~\ref{f:depth}), and represents the total thermal and instrument resolution
broadening. Similar observations of relatively flat
DOS near $E_{\rm F}$ have been made using bulk-sensitive
low-energy laser-PES
at 7~eV.\cite{eguchi2006} In those measurements, performed with much higher
resolution than is achievable at hard x-ray energies, a distinct drop-off of
the QP
DOS within 100~meV of $E_{\rm F}$ is observed in the extracted
bulk component. Although we are unable to resolve such fine details, our
observation of flat DOS near $E_{\rm F}$, coupled with the metallic nature of
these spectra,
demonstrate that recoil effects during the photoemission process at
these energies are minor, and do not significantly shift the spectra, unlike
other materials.\cite{suga2009} In addition to the broader resolution function,
the integration over several Brillouin zones in the current
HAXPES measurements also likely contributes to the flattening of the observed
spectral weight near $E_{\rm F}$.

In order to obtain quantitative information on the spectral weight and
energetics of the LHB, the spectra in Fig.~\ref{f:vb3d} have been fitted to a
linear combination of three Gaussian components, with the results multiplied by
the effective Fermi function. The results of this fit, which describe the data
very well, are shown by the black
solid lines in Fig.~\ref{f:vb3d}, and the dashed line indicates the component
due to the tails of the O~$2p$ states. The other two components represent
photoelectrons from the LHB and QP respectively.

\begin{figure}[t!]
\begin{center}
\includegraphics[width=0.8\linewidth,clip]{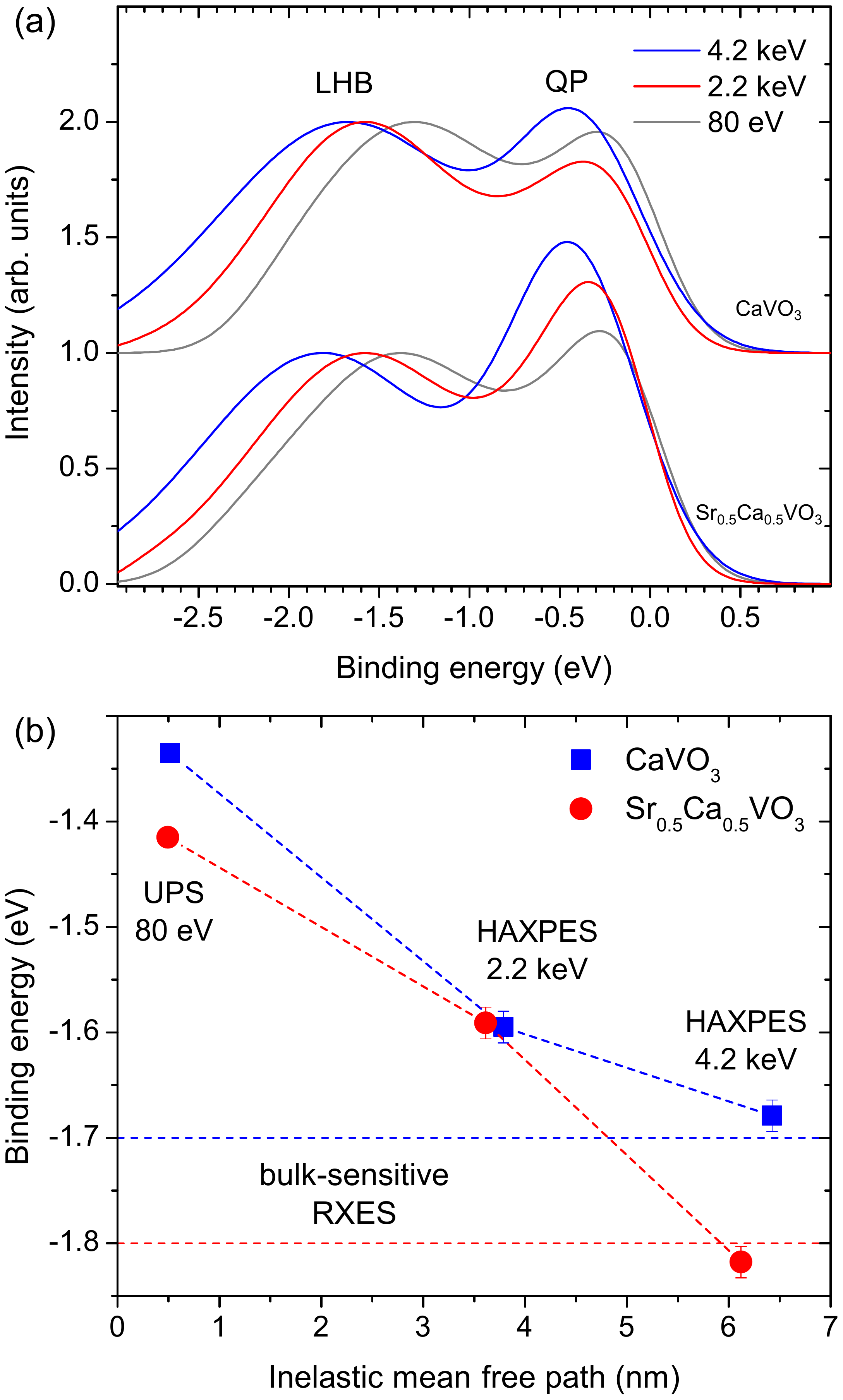}
\end{center}
\vspace*{-0.2in}
\caption{\label{f:lhb} (Color online) Results of fitting the HAXPES spectra
to Gaussian components.
(a)~Fitted spectra (excluding the O $2p$ tails) shown alongside the UPS spectra
of Ref.~\onlinecite{laverock2013c}. All spectra have been broadened to have the
same effective instrument resolution function, corresponding to $\sigma =
220$~meV, and normalised to the intensity of the LHB.
(b)~Binding energy of the LHB for the two
compounds, shown as a function of the estimated IMFP, $\lambda$. The
horizontal dashed lines indicate the separation between the QP
and LHB from bulk-sensitive RXES.\cite{laverock2013c} The error bars represent
estimated statistical errors associated with the fitting.}
\end{figure}

\subsection{Spectral weight}
At both excitation energies shown in Fig.~\ref{f:vb3d}, the relative intensity
of the QP is found to be smaller in CVO than it is in SCVO, which is consistent
with the common notion that CVO is ``more correlated''. In CVO, the narrower
bandwidth, $W$, of the V~$3d$ states means the relative importance of electron
correlations, often parameterized through $U/W$, is greater (the Coulomb energy,
$U$, is not expected to evolve between SrVO$_3$ and
CaVO$_3$\cite{nekrasov2005}).

In order to investigate how the spectral weight varies with HAXPES excitation
energy, we show the fitted spectra normalized to the intensity of the LHB in
Fig.~\ref{f:lhb}(a), alongside the UPS results of
Ref.~\onlinecite{laverock2013c}. To aid comparison, all spectra in
Fig.~\ref{f:lhb}(a) have been broadened to have the same instrument resolution
function of $\sigma = 220$~meV (i.e.~the resolution of the 4.2~keV
HAXPES measurement). Figure~\ref{f:lhb}(a) illustrates that the QP spectral
weight is suppressed in the more surface-sensitive 2.2~keV HAXPES measurement.
Since both the QP and LHB features derive from the same
V~$3d$ states, their photoionization cross-section is the same, and therefore
this evolution can be directly associated with the different depth sensitivities
of the two measurements. The UPS spectra do not follow this trend; however,
these spectra were recorded on oriented samples near the $\Gamma$-point of the
Brillouin zone, and
therefore represent a quite different (and specific) subset of momentum space.

Similar suppression in the QP weight with increasing
surface sensitivity has previously been reported at lower photon
energies.\cite{sekiyama2004,maiti2001,maiti2006}
This observation suggests that there is an {\em intrinsic} surface component of
the electronic structure that is significantly ``more correlated'' than the bulk
in both materials. Moreover, it indicates that an excitation energy of
2.2~keV is not sufficient to
represent the bulk electronic structure of these materials. Previous efforts to
extract bulk information from Sr$_x$Ca$_{1-x}$VO$_3$ have focused on PES at
soft x-ray energies up to 900~eV\cite{sekiyama2004}, 275~eV\cite{maiti2006} or
1486~eV\cite{maiti2001} (although
bulk-sensitive laser-PES has also been used\cite{eguchi2006} the LHB lies
outside the experimental range). These have yielded conflicting
pictures of the bulk electronic structure, for example whether it is
independent of $x$ or not. Our measurements illustrate that hard x-rays
above $\sim 4$~keV are required to truly examine the bulk using photoemission.

\subsection{Binding energy}
Whereas changes in the spectral weight of the QP and LHB are useful indications
of correlated electron behavior, extracting such quantitative information from
experiments
can be complicated by overlapping extrinsic surface
species and the tails of the O~$2p$ states, as well as
the details of the Brillouin zone coverage.
We now turn to the {\em energetics} of the LHB, which are experimentally much
more robust, particularly since the LHB is not dispersive in
momentum.\cite{nekrasov2006,takizawa2009} As can be seen
in Figs.~\ref{f:vb3d} and
\ref{f:lhb}, the energy of the LHB in both compounds lies at deeper energies
in the 4.2~keV spectrum than at 2.2~keV, consistent with stronger electron
correlations at the surface.

At 4.2~keV, the LHB is found at $-1.68$ and $-1.82$~eV for CVO and SCVO
respectively (Fig.~\ref{f:lhb}), which compares very well with the separation
between the QP and LHB observed in truly bulk-sensitive RXES
measurements.\cite{laverock2013c} Qualitatively, these results are in agreement
with dynamical mean-field theory (DMFT)
calculations of CaVO$_3$ and SrVO$_3$ using the same Coulomb
parameter, $U$.\cite{nekrasov2005} HAXPES at this energy can be considered
representative of the bulk correlated electronic structure.  At 2.2~keV the LHBs
of both compounds are located at $-1.59$~eV, which is in rough agreement with
the $x$-independent soft x-ray results.\cite{sekiyama2004} We discuss
these spectra in more detail below, including the similarity in the LHB energy
of both compounds. At the surface, UPS
suggests a markedly shallower energy for the LHB of $-1.34$ and $-1.42$~eV
respectively, which is $\sim 20$\% closer to $E_{\rm F}$ than in the bulk. The
evolution in correlated electron behavior between the surface and the bulk has
been investigated by Ishida {\em et al.}\ through LDA+DMFT calculations of the
VO$_2$- and SrO-terminated surfaces of SrVO$_3$.\cite{ishida2006} At the
VO$_2$-terminated surface, the overall effect on the energy of the LHB is
only weak, partly
due to the emptying of the in-plane $d_{xy}$ surface orbital. On the other hand,
in good agreement with our results, these calculations find that the LHB is
located at $-1.6$~eV in the bulk and $-1.2$~eV at the SrO-terminated surface,
and that the LHB has substantially greater spectral weight at the surface. This
reduction in the binding energy of the correlated Hubbard bands, which is
reproduced for both unrelaxed and relaxed SrO-terminated surfaces, arises
intrinsically due to the reduced coordination of the V ions at the surface, and
the subsequent narrowing of the bandwidth of the out-of-plane $d_{xz,yz}$
orbitals.\cite{ishida2006}

\subsection{2.2 keV HAXPES}
Having established the variation in the energy and spectral weight of the LHB
with depth, we now briefly discuss some of the implications of the spectra
recorded at 2.2~keV. In particular, the energies of the LHBs of both compounds
are significantly shallower than in the bulk, and are very similar for the two
different compounds, despite their differences at both the surface and the bulk.
If we consider that the 2.2~keV spectra contain distinct surface and bulk
components, it is possible to adequately fit the data using LHB and QP energies
representative of the surface (UPS) and bulk (4.2~keV HAXPES). Under this
assumption, 29\% and 45\% of the 2.2~keV HAXPES signal of CVO and SCVO,
respectively, is found to correspond to the surface signal. Indeed, this simple
model reflects the larger surface sensitivity of identical photoemission
measurements on SrVO$_3$ compared with CaVO$_3$, which is due to the higher
density of SrVO$_3$ and therefore shorter IMFP.

However, the results at 2.2~keV do not agree well with simple calculations of
the expected surface contribution. Based on the IMFP,
$\lambda$, the surface contribution can be expressed as $1-\exp(-s/\lambda)$,
where $s$ is the thickness of the surface layer. Assuming a surface layer of
$7.5$~{\AA},\cite{sekiyama2004} the surface contribution is 18\% and 11\% at
2.2~keV and 4.2~keV respectively, with only a very minor difference ($\sim
0.5$\%) between compounds. This is not only at odds with the results presented
above, but it is also impossible to find a reasonable value of the surface
thickness at which 2.2~keV photoelectrons are significantly (e.g.~more than two
times) more sensitive to than 4.2~keV photoelectrons. It therefore appears that
either the distribution of detected photoelectrons from the sample is not simply
an exponential decay, or that the notion of a definite, discrete correlated
surface layer is incorrect. The second possibility seems less likely based on
LDA and DMFT calculations, in which changes to the electronic (and indeed,
crystal) structure due to
the surface were found to decay very rapidly from the surface.\cite{ishida2006}
In either case, it calls into question one of the
common methods of extracting bulk contributions from PES, at least applied to
studying correlated electron
behavior,\cite{maiti2001,maiti2006,sekiyama2004,eguchi2006} in which a spectrum
at a single photon energy is assumed to be composed of discrete surface and bulk
contributions weighted by the exponential decay due to the IMFP. In our data,
it is not possible to reproduce either surface or bulk spectra by such a method.

\begin{figure}[t!]
\begin{center}
\includegraphics[width=1.0\linewidth,clip]{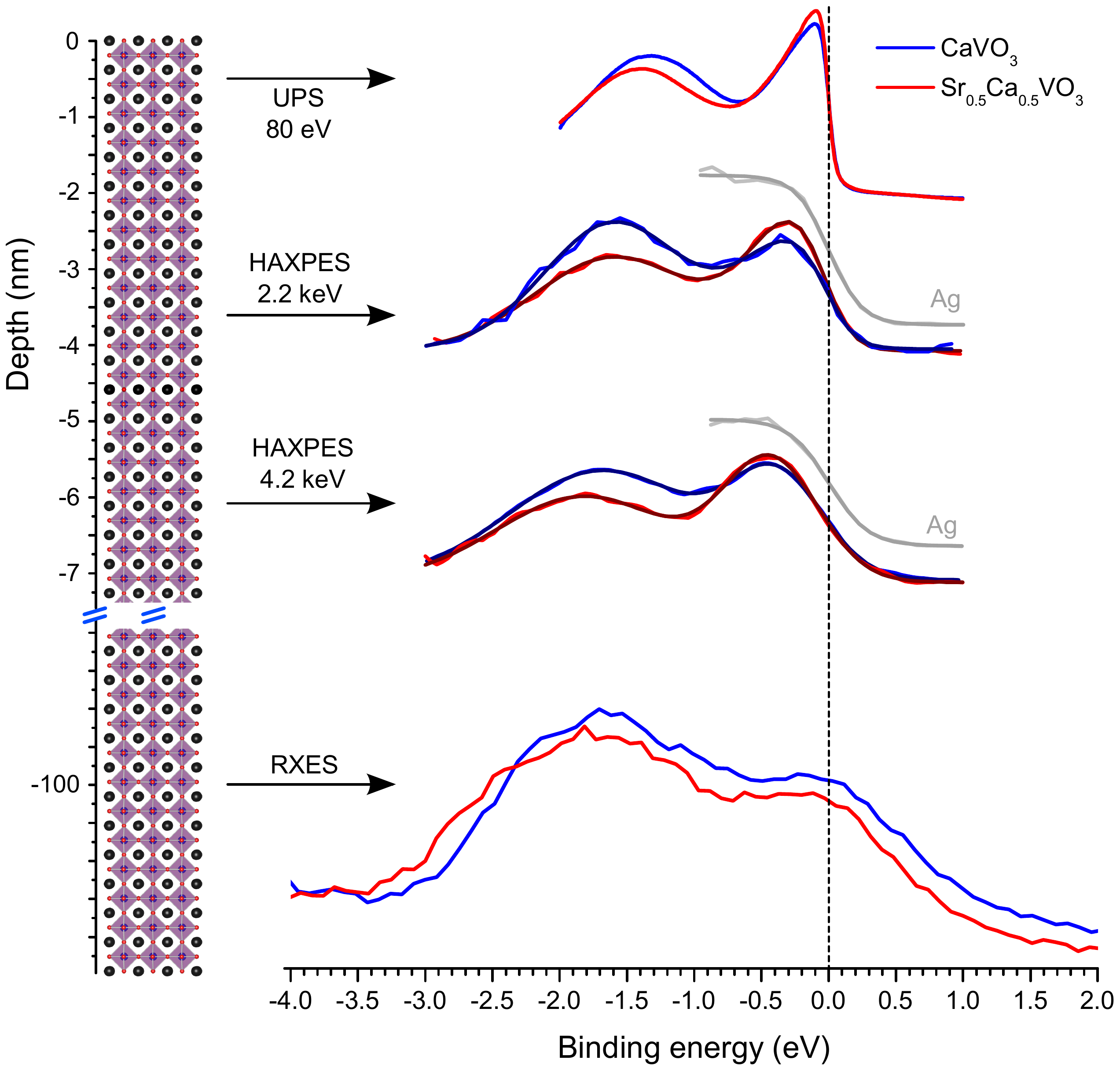}
\end{center}
\vspace*{-0.2in}
\caption{\label{f:depth} (Color online) Comparison between various
probes of different depth sensitivities of the correlated electronic structure
of CaVO$_3$ and Sr$_{0.5}$Ca$_{0.5}$VO$_3$. The
approximate depth sensitivity of each technique is illustrated on the left, with
reference to the SrO-terminated surface of cubic SrVO$_3$(100). The HAXPES
spectra are those shown in Fig.~\ref{f:vb3d} after subtraction of the O $2p$
tails. Corresponding Ag reference spectra recorded at the same energies are also
shown. The UPS and RXES results are from Ref.~\onlinecite{laverock2013c}.}
\end{figure}

\subsection{Summary}
In Fig.~\ref{f:depth}, the O~$2p$ tails have been subtracted from the spectra,
and the results have been compiled with the UPS and RXES measurements
Ref.~\onlinecite{laverock2013c}, which were recorded on the same samples as used
in this study. In this figure, the RXES
spectra, which represent transitions from the valence band into the V~$2p_{3/2}$
core level, have been shifted in energy by 515.2~eV, corresponding to the
coherently-screened V~$2p_{3/2}$ HAXPES core level in Fig.~\ref{f:shcore}.  The
light gray lines show spectra of polycrystalline Ag at 2.2 and 4.2~keV for
comparison, underlying the metallicity of the CaVO$_3$ and
Sr$_{0.5}$Ca$_{0.5}$VO$_3$ measurements. In
Fig.~\ref{f:depth}, the evolution in energy of the LHB from the surface (UPS) to
the bulk (4.2~keV HAXPES) can be clearly seen, as well as the agreement in
energy of the LHB in the bulk HAXPES and the RXES spectra.

In summary, our results illustrate the evolution in correlated electron behavior
from the surface of CaVO$_3$
and Sr$_{0.5}$Ca$_{0.5}$VO$_3$ to the bulk. Spectra at 4.2~keV agree very well
with truly bulk sensitive RXES measurements, and are considered representative
of the bulk correlated electronic structure. On the other hand, spectra at
2.2~keV are in good agreement with those at soft x-ray
energies,\cite{sekiyama2004} but are not bulk-like and still contain a surface
signal of 30 -- 40\%.  We also find that reliable surface and bulk components
cannot be extracted from intermediate spectra, indicating that either the
surface layer is not a discrete overlayer to the bulk, or that detected
photoelectrons do not originate from an exponential distribution within the
sample.

Signatures of evolving electron correlations are observed in both the spectral
weight of the QP and LHB and in the binding energy of the LHB. At both the
surface and in the bulk, CaVO$_3$ is found to be ``more correlated'' than
Sr$_{0.5}$Ca$_{0.5}$VO$_3$: the correlated LHB of CaVO$_3$
exhibits greater spectral weight
and is located closer to $E_{\rm F}$ than in Sr$_{0.5}$Ca$_{0.5}$VO$_3$,
in good agreement with
predictions from DMFT of CaVO$_3$ and SrVO$_3$.\cite{nekrasov2005} At the
surface, the spectral weight of the LHB of both compounds is enhanced, and is
$\sim 20$\% closer to $E_{\rm F}$ than in the bulk, indicating significantly
stronger electron correlations at the surface. This result is in very good
agreement with DMFT calculations of the SrO-terminated SrVO$_3$(001) surface,
which indicate a 25 -- 40\% decrease in binding energy of the LHB at the
surface.\cite{ishida2006}

\section{Conclusion}
We have presented a detailed HAXPES study of the electronic structure of
Sr$_x$Ca$_{1-x}$VO$_3$. Measurements of the valence band have been
quantitatively compared with {\em ab initio} calculations within the LDA, and
are found to be in good agreement after some modest empirical adjustments to the
orbital-dependent photoionization cross-sections, with the $A$-site
hybridization in the O~$2p$ manifold mostly responsible for the differences
between compounds. Secondly, our results on the correlated V~$3d$ bands support
CaVO$_3$ as being a ``more correlated'' metal than SrVO$_3$, and illustrate the
significant enhancement in electron correlations at the surface compared with
the bulk. The results are found to be in good agreement with DMFT predictions of
the evolution in the spectral function with $x$ and between the surface and the
bulk. Finally, we resolve the long-standing controversy of efforts to extract
bulk information from PES by showing that photoelectrons of $\gtrsim 4$~keV are
required to obtain truly bulk-representative spectra.

\section*{Acknowledgements}
The Boston University program is supported in part by
the Department of Energy under Grant No.\ DE-FG02-98ER45680. Supported also
in part by the Boston University/University of Warwick collaboration fund.
The National Synchrotron Light Source, Brookhaven, is supported by the US
Department of Energy under Contract No.\ DE-AC02-98CH10886. Additional support
was provided by the National Institute of Standards and Technology.
G.B.\ gratefully
acknowledges financial support from EPSRC Grant I007210/1.

\end{document}